\documentstyle[11pt,newpasp,twoside,graphics]{article}
\begin{document}
\title{Radio Microlensing: Past, Present \& Near Future}
 \author{L.V.E. Koopmans$^{1,2,3}$, A.G. de Bruyn$^{2,4}$, C.D. Fassnacht$^5$, 
J.~Wambsganss$^6$ \& R.D. Blandford$^1$ + CLASS collaboration}
\affil{$^1$Caltech, mail code 130--33, Pasadena 91125, US\\
$^2$Kapteyn Institute, P.O.Box 800, NL-9700 AV Groningen, Netherlands\\
$^3$JBO, Lower Withington,
Macclesfield, Cheshire SK11 9DL, UK\\
$^4$NFRA-ASTRON, P.O.Box 2, NL-7990 AA Dwingeloo, Netherlands\\
$^5$NRAO, P.O.Box 0, Socorro, NM 87801, USA\\
$^6$U. of Potsdam, Inst. for Phys.,
Am Neuen Palais 10, 14469 Potsdam, Germany}

\begin{abstract}
Strongly correlated {\sl non-intrinsic variability} between 5 and
8.5\,GHz has been observed in one of the lensed images of the
gravitational lens B1600+434.  These non-intrinsic (i.e. `external')
variations are interpreted as {\sl radio-microlensing} of relativistic
$\mu$as-scale jet components in the source at a redshift of $z$=1.59
by massive compact objects in the halo of the edge-on disk lens galaxy
at $z$=0.41. We shortly summarize these observations and discuss
several new observational and theoretical programs to investigate this
new phenomenon in more detail.
\end{abstract}

\vspace{-0.5cm}

\section{Radio-Microlensing: B1600+434}

Multi-frequency observations of the CLASS gravitational lens (GL) B1600+434
with the VLA and WSRT radio telescopes at 1.4, 5 and 8.5\,GHz has
unambiguously shown {\sl non-intrinsic variability} in the lensed image
that passes through the dark-matter halo of the edge-on disk lens
galaxy (Koopmans \& de Bruyn 2000; Koopmans et al. 2000a). Based on the
amplitude--timescale and frequency dependence of these non-intrinsic
variations {\sl and} the difference in variability between the two
lensed images, separated by only 1\farcs4, the {\sl non-intrinsic
variability} is interpreted as {\sl radio-microlensing} of
relativistic $\mu$as-scale jet components in the source by massive
compact objects in the halo of the edge-on disk lens galaxy. The
alternatives, i.e.  scintillation and extreme scattering, have
considerable difficulties in explaining the time-scale and frequency
dependence of the observations (e.g. Koopmans \& de Bruyn 2000;
Koopmans et al. 2000a). An example of these
variations at 5\,GHz is shown in Fig.1. The strong non-intrinsic event
between days 70--100 is interpreted as a {\sl radio-microlensing
caustic crossing}. This type of event is extremely difficult to
explain in terms of scattering (ISS) by the ionized interstellar medium
(e.g. Koopmans et al. 2000b).

\section{Radio-Microlensing: Other Lens Systems \& Future }

To investigate {\sl radio-microlensing} in B1600+434 and other lens
systems in more detail, we have started a number of new monitoring
programs with the WSRT, VLA and MERLIN radio telescopes. With the WSRT
and VLA, we have been and are currently monitoring B1600+434 at 1.4,
(1.7), 5.0, 8.5, 15 and (22)\,GHz. (Wavelengths between
parentheses cover only part of the light curves.) This will enable us
to further constrain the microlensing and ISS hypotheses based on
their frequency dependence. In addition, we are currently monitoring 8
CLASS GL systems with the VLA at 8.5\,GHz to measure additional
time-delays for the determination of H$_0$. These lightcurves,
however, also provide information on {\sl radio-microlensing}. In
addition, a MERLIN Key--Project to monitor 12 CLASS GL systems (2/3
quads) at 5\,GHz will conditionally start around Dec. 2000/Jan. 2001,
with the aim of finding new cases of {\sl radio-microlensing}. Combined, 
we will be able to search for this phenomenon in 14 aditional radio GL systems.

Besides this ongoing observational effort, we are theoretically
studying {\sl radio-microlensing} in B1600+434, using numerical
microlensing simulations, to place constraints on the fraction and
mass-function of massive compact objects in the lens galaxy (Koopmans
\& Wambsganss, in prep).

\begin{figure*}[t!]
\resizebox{6.5cm}{!}{\includegraphics{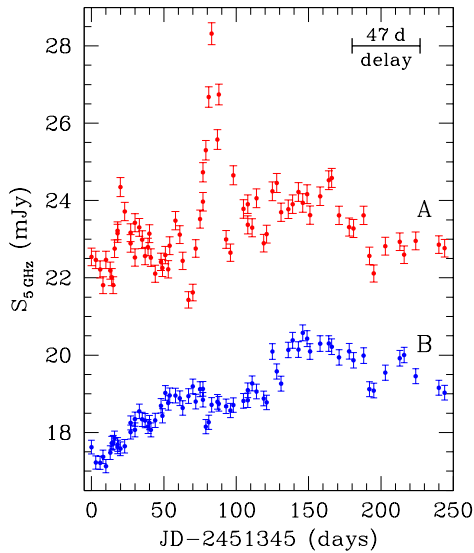}}
\parbox[b]{6.5cm}{
\caption{Preliminary results (at 5\,GHz) from the 1999/2000 VLA
monitoring campaign of B1600+434. The upper light curve~(image A)
passes through the dark-matter halo of the edge-on spiral lens galaxy.
Note several strong (up to 30\%) events in the upper lightcurve and
the complete absence of these events in the lower light curve (image
B) after the measured time delay of $\sim$47 days.}\vspace{0.8cm}}
\end{figure*}

\end{document}